\documentclass[a4paper,11pt]{article}

\usepackage{pos}
\usepackage{subcaption}
\usepackage{lipsum} 
\usepackage{multicol}
\usepackage{wrapfig}

\title{Enhancing searches for astrophysical neutrino sources in IceCube with machine learning and improved spatial modeling}

\ShortTitle{Improved spatial modeling and PeVatron searches in IceCube}

\author{The IceCube Collaboration \\{\normalsize \normalfont(a complete list of authors can be found at the end of the proceedings)}\\}

\emailAdd{lseen@icecube.wisc.edu}
\emailAdd{tyuan@icecube.wisc.edu}
\emailAdd{lu.lu@icecube.wisc.edu}
\emailAdd{thiesmeyer@icecube.wisc.edu}
\emailAdd{karle@icecube.wisc.edu}

\abstract{

Searches for astrophysical neutrino sources in IceCube rely on an unbinned likelihood that consists of an energy and spatial component. Accurate modeling of the detector, ice, and spatial distributions leads to improved directional and energy reconstructions, resulting in increased sensitivity. In this work, we utilize our best knowledge of the detector ice properties and detector calibrations to reconstruct in-ice particle showers. The spatial component of the likelihood is parameterized either by a 2D Gaussian or a von Mises Fisher (vMF) distribution at small and large angular uncertainties, respectively. Here, we use a gradient-boosted decision tree with a vMF spatial likelihood loss function, reparameterized through two coordinate transformations, to predict per-event point spread functions (PSF). Additionally, we discuss the search for PeV cosmic ray sources using the IceCube Multi-Flavor Astrophysical Neutrino (ICEMAN) sample. Our search contains both an analysis of individual neutrino sources coincident with greater than 100 TeV gamma-ray sources and also a stacking analysis. We outline the prospects for extended neutrino emission originating from the Cygnus Cocoon region.

\vspace{4mm}

{\bfseries Corresponding authors:}
Leo Seen$^{1*}$, 
Tianlu Yuan $^{1}$, 
Lu Lu $^{1}$, 
Matthias Thiesmeyer$^{1}$,
Albrecht Karle$^{1}$\\
{$^{1}$ \itshape Dept. of Physics and Wisconsin IceCube Particle Astrophysics Center, University of Wisconsin}\\[4mm]
$^*$ Presenter
}

\ConferenceLogo{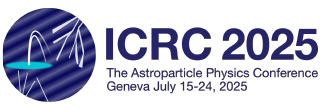}

\FullConference{39th International Cosmic Ray Conference (ICRC2025)\\
 15–24 July 2025\\
Geneva, Switzerland\\}

\begin{document}

\maketitle

\section{Introduction}\label{sec1}
The location and nature of neutrino sources remain mostly unknown. Discovering these sources will advance our understanding of astrophysical neutrino production, cosmic ray acceleration, and the origin of cosmic rays. To date, evidence for neutrino emission from point sources has only been observed in the blazar TXS 0506+056 \cite{txs_neutrino_source} and active galaxy NGC 1068 \cite{ngc_1068} at 3.5 and 4.2 sigma respectively. More recently, using ten years of data and a neural network-based in-ice particle shower dataset (DNN Cascades), IceCube reported extended neutrino emission originating from the galactic plane \cite{galactic_plane} at 4.5 sigma. However, whether these neutrinos originate from diffuse emission or point sources remains unclear. Improving IceCube's sensitivity to both point and extended sources will require improved reconstruction methods, an updated understanding of detector ice properties, and refined angular error estimators. 

\begin{wrapfigure}{r}{0.5\textwidth}
    \centering
    \vspace{-20pt}
    \includegraphics[width=0.5\textwidth]{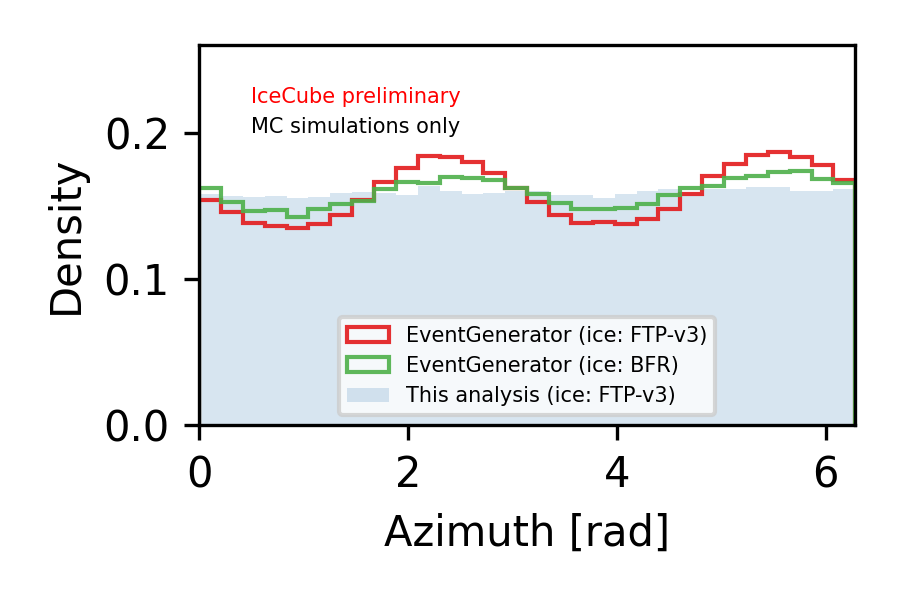}
    \caption{Normalized azimuth distributions for neutrino-induced cascades from MC simulations only. Solid lines compare the EventGenerator predictions using the SPICE FTP-v3 ice model (red) and the SPICE BFR ice model (green), while the filled histogram (light blue) shows the same SPICE FTP-v3 prediction processed through the PreferredFit reconstruction. The expected azimuth distribution is flat.}
    \label{fig:ice_modeling}
\end{wrapfigure}

In the DNN Cascades dataset, a neural network predicts per-event PSFs. Recent developments in machine learning have shown that tree-based models outperform neural networks when trained on tabular data \cite{grinsztajn2022treebasedmodelsoutperformdeep}. Reasons for this outperformance are still debated and remain an active area of research. An example of a tree-based model is the Gradient Boosted Decision Tree (GBDT) \cite{gbdt_original}. GBDTs provide a fast and customizable framework that can perform regression and classification tasks. Specifically, the ability to implement custom loss functions and scoring metrics allows us to align them with the log-likelihood \cite{Braun_2008} used in IceCube neutrino source analyses. 

In this work, we present the results of a GBDT model, trained using the lightgbm software package \cite{lightgbm}, to predict per-event PSFs for an updated DNN Cascades dataset. Additionally, we show the effects of improved reconstruction and ice modeling. We then combine this updated DNN Cascades dataset with starting tracks and through-going muons to create a 12.3-year ICEMAN sample. Finally, we will discuss the applications of the ICEMAN sample in the context of galactic PeVatrons, hypothesized sources within the Milky Way capable of accelerating cosmic rays up to PeV energies. 

\section{Ice Modeling/Reconstruction Methods}
Accurate modeling of the ice leads to significantly improved event reconstruction and angular resolutions \cite{Abbasi_2024}. We utilize the South Pole ICE (SPICE) Full Tilt Parameterization (FTP-v3) model \cite{chirkin2023improvedmappingicelayer}, our best understanding of the detector ice. Furthermore, we use updated methods of reconstructing in-ice particle showers, Taupede and Monopod \cite{Abbasi_2024}. Monopod performs a maximum likelihood fit assuming a single cascade, while Taupede assumes a double cascade event morphology. Notably, Taupede uses the results from Monopod as seeds in its fit. Both reconstruction methods are applied per event, with the better fit chosen as the PreferredFit. Biases in the azimuth distribution originate from ice anisotropies and ice flows. By using the SPICE FTP-v3 ice model and the PreferredFit reconstruction, biases in azimuth decreases, as shown in Figure \ref{fig:ice_modeling}. 

\section{GBDT Training Inputs}
From the PreferredFit reconstruction, we select 14 reconstructed quantities, listed in Table \ref{tab:input_quantities}, to use as input features into the GBDT. 

\begin{table}[h!]
\centering
\begin{tabular}{|l|l|l|}
\hline
Reco. interaction vertex $x,y,z$ & Total deposited charge & Reco. energy\\
\hline
Reco. zenith and azimuth & ndof  & Length between two cascades\\
\hline 
Distance from reco. vertex & Shortest distance from reco. & Reco. qualities: rlogl,\\
to closest string & vertex to detector edge & square residual, chi-squared\\
\hline

\end{tabular}
\caption{List of input parameters into the GBDT. Reco is short for Reconstructed/Reconstruction. The length between two cascades is set to zero if a single cascade is preferred. ndof stands for number of degrees of freedom, a proxy for the number of digital optical modules used in the fit.}\label{tab:input_quantities}
\end{table}

Our Monte-Carlo (MC) simulation dataset before final cuts, totals 5,718,881 events, 2,570,907 $\nu_\mu$, 1,737,306 $\nu_\tau$, and 1,410,668 $\nu_e$ events. A final level cut on muons will reduce the number of $\nu_\mu$ and $\nu_\tau$ events. Each set of neutrino flavors has events simulated at three separate energies: 100 GeV to 10 TeV, 10 TeV to 1 PeV, and 1 PeV to 100 PeV. We use 50\% of each neutrino flavor in each energy bin for training and the other 50\% for testing. After selecting the training set, we weigh each event proportional to an $E^{-2.7}$ spectrum.

\section{Likelihood/Loss Function Construction}
IceCube utilizes an unbinned maximum likelihood approach \cite{Braun_2008} to search for neutrino sources. The standard likelihood contains a signal and background term, factorized into independent space and energy terms. The per-event PSFs contribute only to the spatial signal component of the likelihood. A vMF, instead of a Gaussian, is used at larger PSFs to accurately characterize the tail of the distribution:
\begin{equation}\label{eqn:psf}
    f(\sigma,\psi) = \begin{cases} \frac{1}{2\pi\sigma^2}e^{-\frac{\psi^2}{2\sigma^2}} & \sigma\leq7^o\\ \frac{1}{4\pi\sigma^2\sinh\left(\frac{1}{\sigma^2}\right)}e^{\frac{\cos(\psi)}{\sigma^2}} & \sigma>7^o\end{cases}.
\end{equation}
Where $\sigma$ is the per-event PSF, and $\psi$ is the opening angle between true and reconstructed directions. In the GBDT framework, the minimum of the loss function after each iteration evaluates the model performance. Additionally, a scoring metric evaluates the final model performance after training. Since we use a maximum likelihood technique to search for neutrino sources, we choose our loss function as the negative log-likelihood. Since the vMF distribution converges to a Gaussian at smaller PSFs, we parameterize our loss function with a vMF distribution across all PSFs, defined as:
\begin{equation}
    L(\sigma,\psi) = -\ln(f(\sigma,\psi)) = \ln(4\pi)+\ln(\sigma^2)+\ln\left[\sinh\left(\frac{1}{\sigma^2}\right)\right]-\frac{\cos(\psi)}{\sigma^2}.
\end{equation}
Since we are minimizing the loss function, neglect constants. Using the coordinate transformations $\kappa = 1/\sigma^2$ and $\eta = \ln(\kappa)$, we arrive at a numerically stable loss function parameterized in terms of $\eta$:
\begin{equation}\label{eqn:loss_function}
    L(\eta,\psi) = e^{\eta} +\ln(1-e^{-2e^{\eta}})-\eta-e^{\eta}\cos(\psi).
\end{equation}
By allowing the GBDT to predict $\eta$, values of $\sigma$ are bounded in the physical range between $(0, \infty)$. Traditionally, lightgbm relies on a hessian-based minimization of the loss function. However, due to numerical instability with the hessian of Equation \ref{eqn:loss_function}, we set the hessian to one and use a gradient-based minimization approach. The gradient follows:
\begin{equation}
    \nabla_\eta(L(\eta,\psi)) = e^{\eta}\coth\left(e^{\eta}\right)-1-e^{\eta}\cos(\psi).
\end{equation}
Furthermore, the average value of Equation \ref{eqn:loss_function} over all events is used as the scoring metric. 

\section{Hyperparameter Optimization}\label{sec x}

Hyperparameters can have significant impacts on a model's performance. To optimize hyperparameters, we conduct a partial grid search over four hyperparameters and select the model yielding the lowest score. Table \ref{tab:hyperparameters} outlines the hyperparameters and values tested. A full grid scan wasn't possible due to computational limitations. While we test models with 1000 and 5000 maximum iterations, the early stopping round is fixed to 50 to prevent overtraining, meaning not all models reach the maximum number of allowed iterations.

\begin{table}[h!]
\centering
\begin{tabular}{lcc}
\hline
Hyper-Parameter  &	Range & Optimized value\\
\hline
\# of leaves & [25, 50, 75, 100, 125, 150] & 100\\
\hline
Minimum data in leaf & [500, 1000, 1250, 1500, 1750, 2000, 5000]& 1750\\
\hline 
Learning Rate & [1e-3, 5e-3, 7.5e-3, 0.01, 0.025, & 0.05\\ 
& 0.05, 0.075, 0.1, 0.15, 0.2] & \\
\hline
\# of iterations & [1000, 5000]& 1000\\

\end{tabular}
\caption{The first column lists the hyperparameters considered in the optimization. The second column lists the values of each parameter that were tested. The third column shows the optimal hyperparameter value found.}\label{tab:hyperparameters}
\end{table}

\begin{wrapfigure}{R}{0.5\textwidth}
    \centering
    \includegraphics[width=0.5\textwidth]{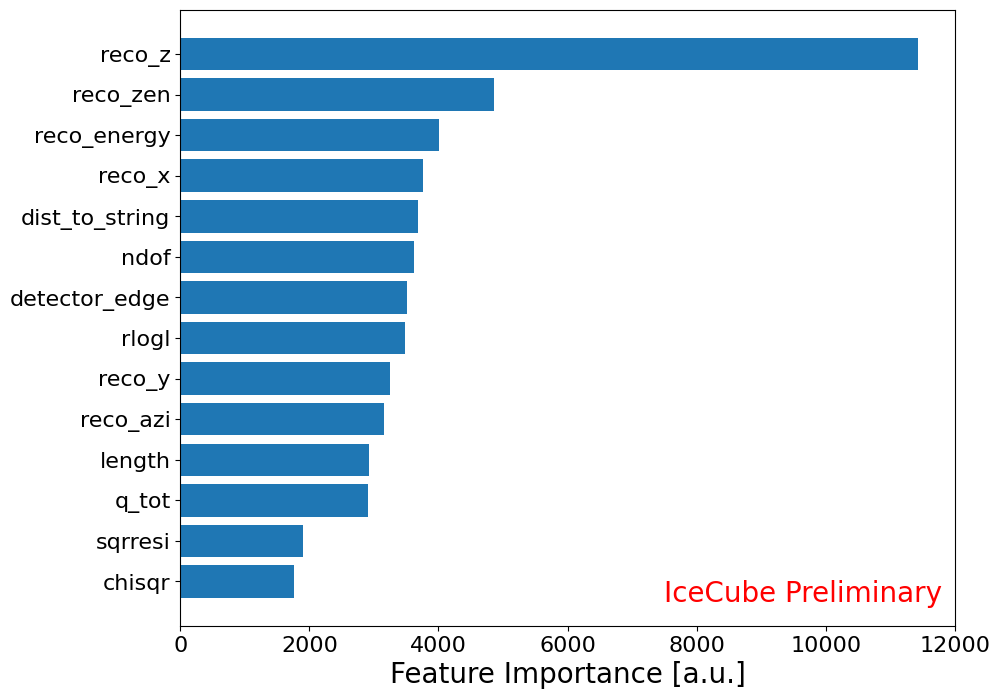}
    \caption{Importance of the 14 input features into the GBDT.}
    \vspace{-30pt}
    \label{feature_importance}
\end{wrapfigure}

\section{GBDT Performance}\label{sec3}

After hyperparameter optimization and training our best model, we evaluate the performance of our GBDT on test data. The final model reaches the early stopping threshold after 548 iterations. After the first 100 iterations, the score of the test set only marginally improves but doesn't worsen, signaling that the model isn't overtrained. From Figure \ref{feature_importance}, we find that the reconstructed depth parameter is the highest-ranked variable. A significant worsening in the PSF in the region around $-200\text{ m}<z<0\text{ m}$ due to the existence of the dust layer \cite{tc-18-75-2024} suggests that the GBDT was able to learn the detector's ice properties. By learning the detector ice properties, the GBDT should more accurately predict the per-event PSFs. 

The third most important variable is the reconstructed neutrino energy. We expect smaller PSFs at higher energies due to more light deposition within the detector. The best median PSF we achieve is 5.97 degrees in the 290 TeV to 13.8 PeV energy range. Above 13.8 PeV, there is a slight worsening in the PSF which could be due to saturation or the Landau–Pomeranchuk–Migdal effect \cite{lpm}. The reason for this worsening is still under investigation.

\begin{figure}[b!]
    \centering
    \begin{subfigure}[t]{0.45\textwidth}
        \centering
        \includegraphics[height=1.8in]{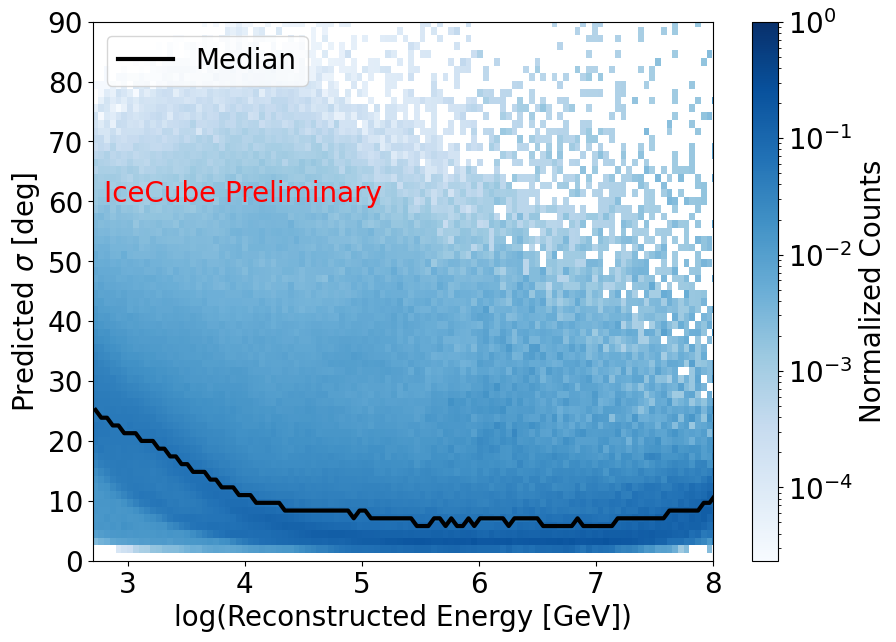}
        \caption{}\label{sigma_vs_logE}
    \end{subfigure}
    ~
    \begin{subfigure}[t]{0.45\textwidth}
        \centering
        \includegraphics[height=1.8in]{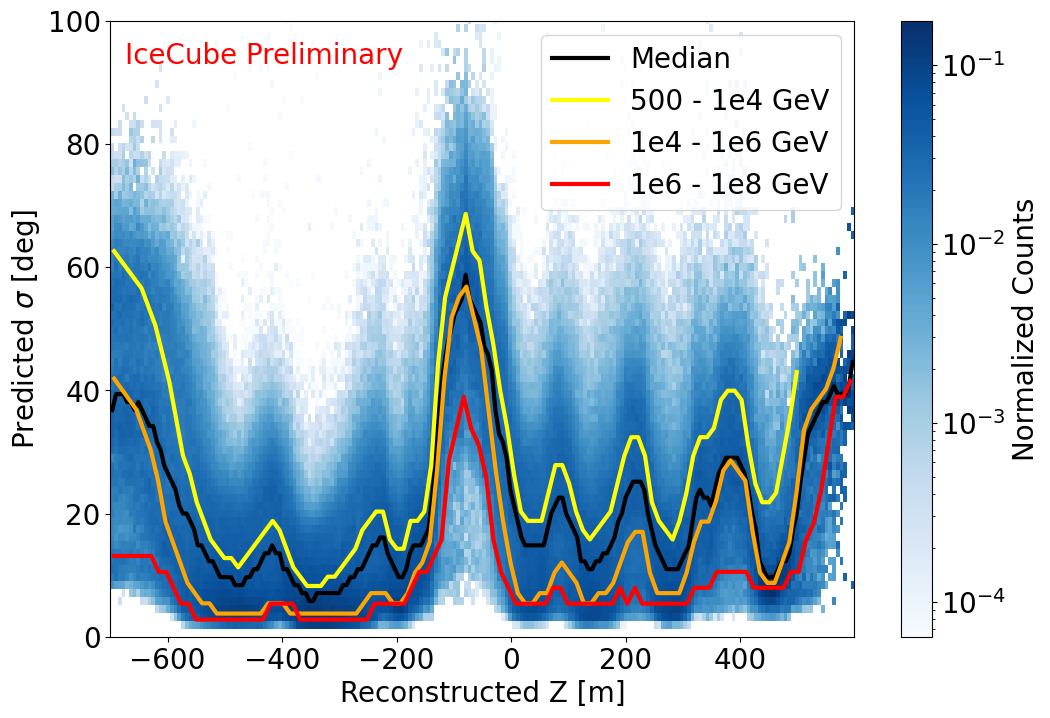}
        \caption{}\label{sigma_vs_z}
    \end{subfigure}
    \caption{Column normalized, 2D histogram of the predicted angular error vs log$_{10}$(reconstructed energy) and reconstructed depth for figures (a) and (b) respectively. Reconstructed energy is in units of GeV while reconstructed depth is in meters. The black solid line shows the median counts in each column.}\label{gbdt_performance_metrics}
\end{figure}

\section{DNN Cascades ICEMAN}\label{sec4}
\begin{figure}[t!]
    \centering
    \begin{subfigure}[t]{0.45\textwidth}
        \centering
        \includegraphics[height=1.8in]{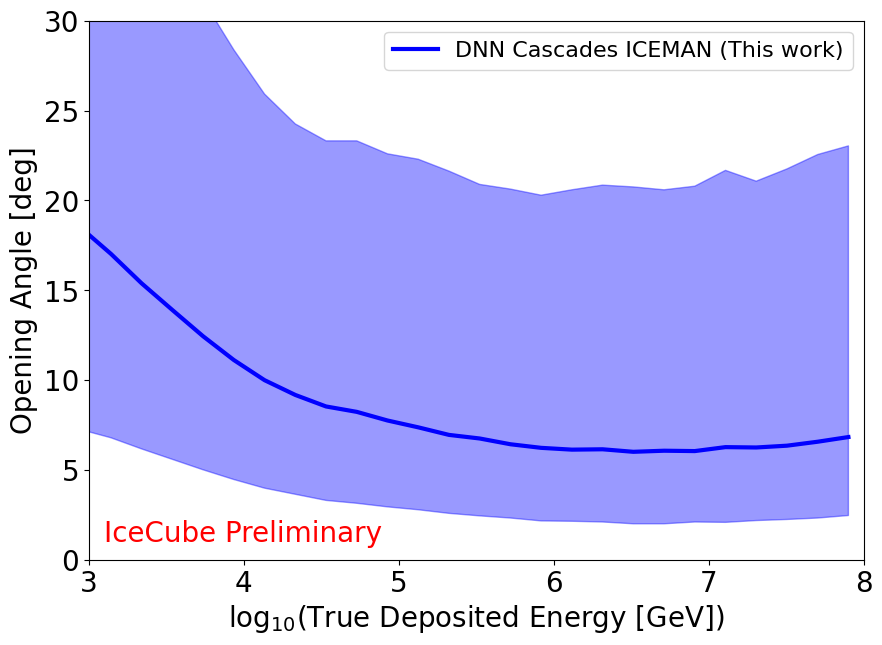}
        \caption{}\label{angular_resolution}
    \end{subfigure}%
    ~ 
    \begin{subfigure}[t]{0.45\textwidth}
        \centering
        \includegraphics[height=1.8in]{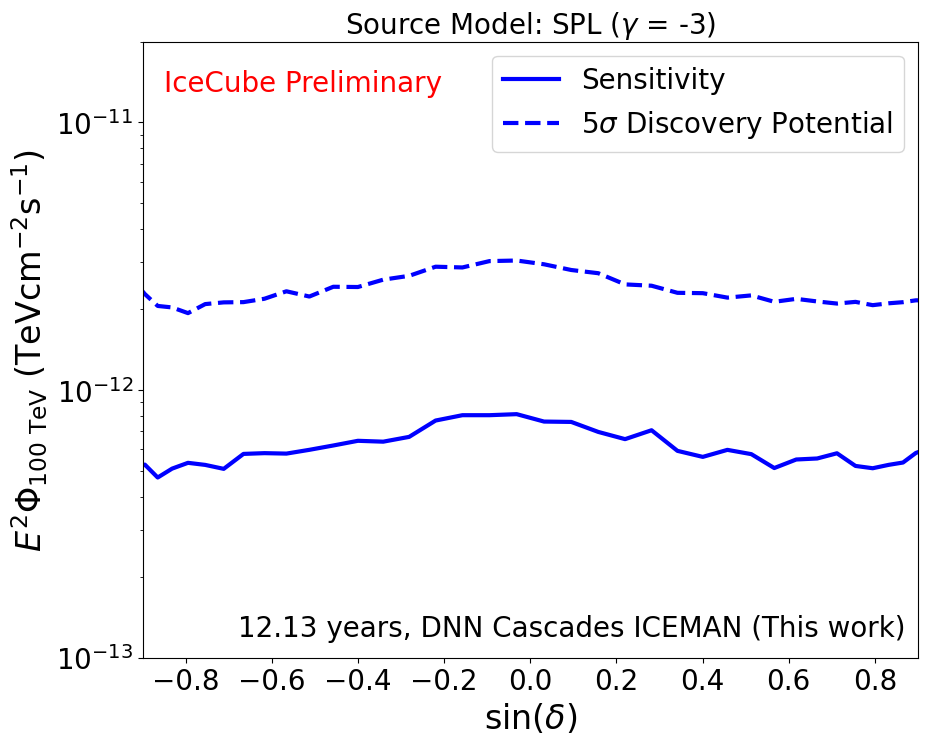}
        \caption{}\label{sensitivity}
    \end{subfigure}
    \caption{Performance of DNN Cascades ICEMAN. (a) The opening angle between true and reconstructed directions as a function of the log(True deposited energy [GeV]). The solid line represent the median, while the shaded portion indicates the 1$\sigma$ containment region. (b) Point source sensitivity and discovery potential calculated assuming a $E^{-3}$ single power law (SPL) spectrum. The solid and dashed lines show the point source sensitivity and discovery potential, respectively.}\label{}
\end{figure}

Applying the same cuts as the previous selection \cite{steve} but removing events with greater than 40-degree PSFs based on the new PSF predictions, we create DNN Cascades ICEMAN. With updated energy reconstructions, in the same 10 years of data, we have 12.35\% more events compared to the original DNN Cascades sample. The smallest median angular resolution is six degrees, as shown in Figure \ref{angular_resolution}. Figure \ref{sensitivity} indicates that with two additional years of data, we obtain a sensitivity and discovery potential on the order of $10^{-13}/10^{-12}$ TeV cm$^{-2}$ s$^{-1}$ assuming a $E^{-3}$ SPL spectrum at 100 TeV.

\section{Future Prospects}
Numerous galactic PeVatron candidates have arisen in recent years. The Large High Altitude Air Shower Observatory (LHAASO), High Altitude Water Cherenkov (HAWC) Observatory, and High Energy Stereoscopic System (HESS) have detected galactic source photon emission extending beyond 100 TeV (\cite{Cao_2024}, \cite{Abeysekara_2020}, \cite{Abdalla_2021}). At these energies, inverse-Compton scattering, the primary method of leptonic production, is greatly suppressed due to the Klein-Nishina effect \cite{Klein1929}. Furthermore, LHAASO has not only reported photon emission from 12 galactic sources with energies up to 1.4 PeV \cite{Cao2021} but also provides the largest greater than 100 TeV gamma-ray catalog to date \cite{lhaaso_catalog}. 

\begin{figure}[h!]
    \centering
    \includegraphics[height=1.8in]{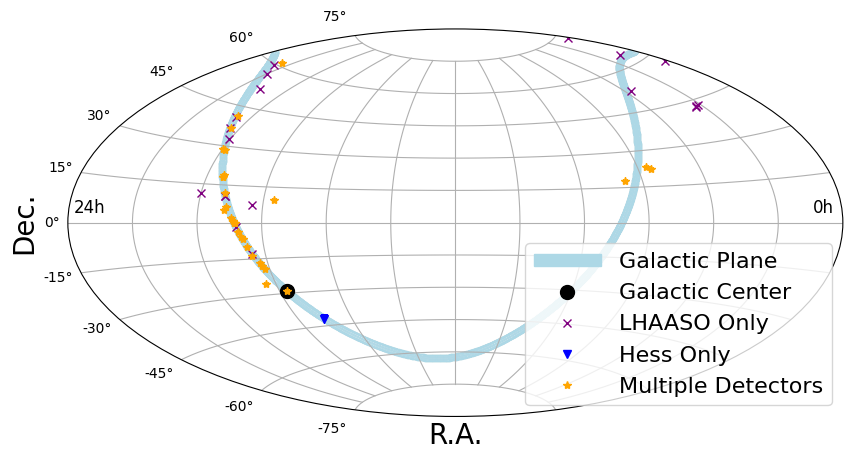}
    \caption{Skymap of all gamma-ray sources with photon emissions above 100 TeV in equatorial coordinates. The light blue band shows the galactic plane. }\label{skymap}
\end{figure}

Assuming some of these photons originate from hadronic interactions, we expect to see a corresponding neutrino flux. We create a catalog of all known gamma-ray sources with photon emission exceeding 100 TeV. Figure \ref{skymap} shows nearly all the sources are located along the galactic plane. First, we will conduct an individual source search of a selected subset of the catalog. The selected sources are those most likely to be hadronic. Then, we will perform a stacking analysis by grouping sources into five source classes - supernova-remnant (SNR), pulsar wind nebula (PWN), microquasars, unidentified (UNID), and pulsars with molecular clouds or SNRs nearby. 

\begin{figure}[t!]
    \centering
    \begin{subfigure}[t]{0.32\textwidth}
        \centering
        \includegraphics[height=1.675in]{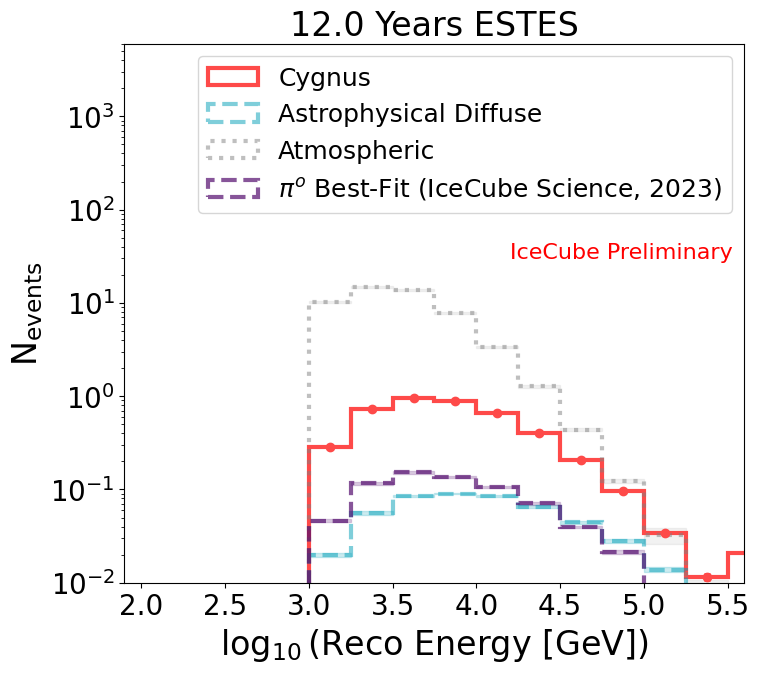}
        \caption{ESTES}\label{cygnus_estes_fermi}
    \end{subfigure}%
    ~ 
    \begin{subfigure}[t]{0.32\textwidth}
        \centering
        \captionsetup{justification=centering}
        \includegraphics[height=1.675in]{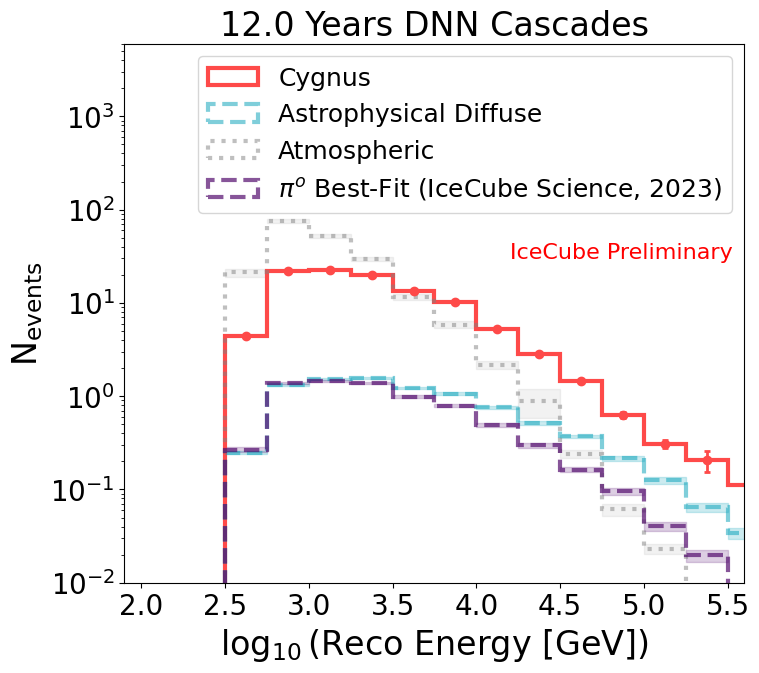}
        \caption{DNN Cascades\\ (IceCube Science, 2023)}\label{cygnus_dnn_fermi}
    \end{subfigure}
    ~
    \begin{subfigure}[t]{0.32\textwidth}
        \centering
        \includegraphics[height=1.675in]{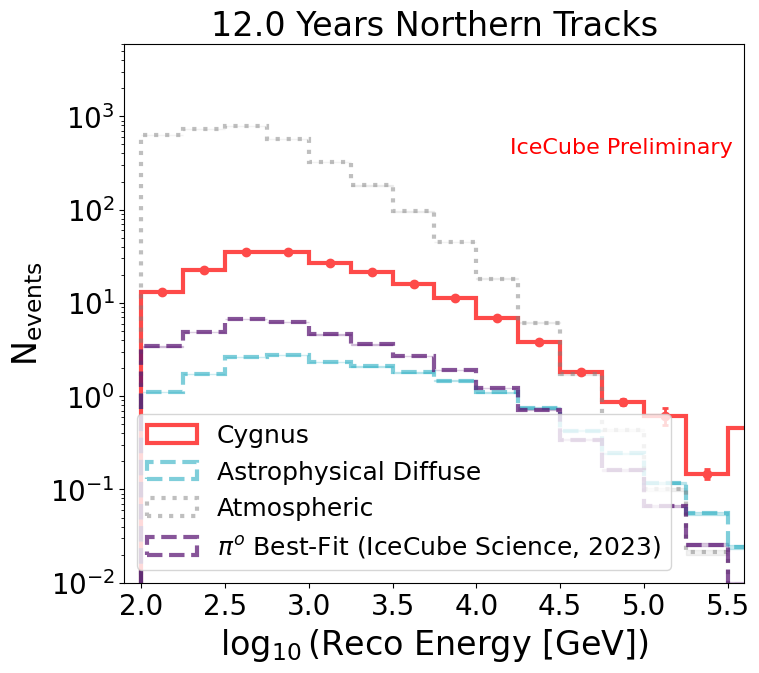}
        \caption{Northern Tracks}\label{cygnus_nt_fermi}
    \end{subfigure}
    \caption{Expected number of events per dataset. The livetimes of each dataset is scaled up to 12 years. The grey dotted line shows the expected number of atmospheric neutrinos derived using the SIBYLL 2.3C hadronic interaction model and the Gaisser H3a cosmic ray model. The red solid line is the number of expected events originating from a 6 degree bubble. The neutrino flux for Cygnus is derived assuming 100\% pp-interaction from the LHAASO spectrum \cite{cygnus_superbubble}. The purple dashed line is the expected neutrino events from galactic diffuse emission, assuming the Fermi-$\pi^0$ best fit model flux from IceCube, Science 2023 \cite{galactic_plane}. The light blue dashed line is the astrophysical diffuse spectrum assuming a single power law of the form, $dN/dE = N_o(E/E_o)^{-\gamma}$ where $N_o = 1.44\times 10^{-8}$ GeV$^{-1}$cm$^{-2}$s$^{-1}$sr$^{-1}$, $E_o = 10^{5}$ GeV, and $\gamma = 2.37$}\label{Cygnus_Cocoon_ER}
    \vspace{-15pt}
\end{figure}

A specific source of interest is the complex Cygnus Cocoon region. LHAASO has reported a gamma-ray super-bubble spanning around 6 degrees \cite{cygnus_superbubble} while HAWC has reported a smaller, two-degree extension \cite{HAWC_cygnus}. The existence of molecular clouds, SNRs, Microquasars, etc. in the region suggests the possibility of hadronic interactions. In Figure \ref{Cygnus_Cocoon_ER}, we show the expected number of events originating from the Cygnus Cocoon region per dataset. For the Enhanced Starting Track Event Selection (ESTES) \cite{ESTES} in Figure \ref{cygnus_estes_fermi}, we expect no Cygnus source neutrino events above the atmospheric background in 12 years of data. While ESTES is a very pure astrophysical neutrino sample, ESTES has much fewer events in total compared to DNN Cascades and Northern Tracks \cite{northern_tracks}. For DNN Cascades, we see in Figure \ref{cygnus_dnn_fermi} that we expect a few events greater than 10 TeV where the Cygnus source contribution dominates the background emission. Due to the worse angular resolution of cascades, localizing events to a specific source is challenging. However, given the spatial extension quoted by LHAASO and HAWC, DNN Cascades could still improve our sensitivity. Finally, in the Northern Tracks sample shown in Figure \ref{cygnus_nt_fermi}, we see that the Cygnus source contribution only begins to dominate the background at energies above 30 TeV. In the northern sky, the Northern Tracks dataset provides most of the point source sensitivity. 

\section{Conclusion}\label{sec5}
In this work, we have shown a flattening in the azimuthal bias from updated reconstruction methods and ice models. Furthermore, we have detailed the construction of a GBDT model trained to predict per-event PSFs utilizing a custom loss function and the resulting DNN Cascades ICEMAN selection. Then, combining DNN Cascades ICEMAN with ESTES and Northern Tracks, we outline the potential of using the combined ICEMAN sample to search for galactic PeVatrons. We plan to conduct an individual point source search for neutrinos coincident with photon sources that emit above 100 TeV and an extended source search of neutrinos originating from the Cygnus Cocoon region. Additionally, we will perform a stacking analysis of these photon sources categorized into various source classes. 

\bibliographystyle{ICRC}

\small{
\bibliography{references}
}

\clearpage

\section*{Full Author List: IceCube Collaboration}

\scriptsize
\noindent
R. Abbasi$^{16}$,
M. Ackermann$^{63}$,
J. Adams$^{17}$,
S. K. Agarwalla$^{39,\: {\rm a}}$,
J. A. Aguilar$^{10}$,
M. Ahlers$^{21}$,
J.M. Alameddine$^{22}$,
S. Ali$^{35}$,
N. M. Amin$^{43}$,
K. Andeen$^{41}$
% ...
\\
\\
% ...
$^{10}$ Universit{\'e} Libre de Bruxelles, Science Faculty CP230, B-1050 Brussels, Belgium \\
% ...
$^{16}$ Department of Physics, Loyola University Chicago, Chicago, IL 60660, USA \\
$^{17}$ Dept. of Physics and Astronomy, University of Canterbury, Private Bag 4800, Christchurch, New Zealand \\
% ...
$^{21}$ Niels Bohr Institute, University of Copenhagen, DK-2100 Copenhagen, Denmark \\
$^{22}$ Dept. of Physics, TU Dortmund University, D-44221 Dortmund, Germany \\
% ...
$^{35}$ Dept. of Physics and Astronomy, University of Kansas, Lawrence, KS 66045, USA \\
% ...
$^{39}$ Dept. of Physics and Wisconsin IceCube Particle Astrophysics Center, University of Wisconsin{\textemdash}Madison, Madison, WI 53706, USA \\
% ...
$^{41}$ Department of Physics, Marquette University, Milwaukee, WI 53201, USA \\
% ...
$^{43}$ Bartol Research Institute and Dept. of Physics and Astronomy, University of Delaware, Newark, DE 19716, USA \\
% ...
$^{63}$ Deutsches Elektronen-Synchrotron DESY, Platanenallee 6, D-15738 Zeuthen, Germany \\
$^{\rm a}$ also at Institute of Physics, Sachivalaya Marg, Sainik School Post, Bhubaneswar 751005, India \\
% ...

\subsection*{Acknowledgments}

\noindent
The authors gratefully acknowledge the support from the following agencies and institutions:
USA {\textendash} U.S. National Science Foundation-Office of Polar Programs,
U.S. National Science Foundation-Physics Division,
U.S. National Science Foundation-EPSCoR,
U.S. National Science Foundation-Office of Advanced Cyberinfrastructure,
Wisconsin Alumni Research Foundation,
Center for High Throughput Computing (CHTC) at the University of Wisconsin{\textendash}Madison,
Open Science Grid (OSG),
Partnership to Advance Throughput Computing (PATh),
Advanced Cyberinfrastructure Coordination Ecosystem: Services {\&} Support (ACCESS),
Frontera and Ranch computing project at the Texas Advanced Computing Center,
U.S. Department of Energy-National Energy Research Scientific Computing Center,
Particle astrophysics research computing center at the University of Maryland,
Institute for Cyber-Enabled Research at Michigan State University,
Astroparticle physics computational facility at Marquette University,
NVIDIA Corporation,
and Google Cloud Platform;
Belgium {\textendash} Funds for Scientific Research (FRS-FNRS and FWO),
FWO Odysseus and Big Science programmes,
and Belgian Federal Science Policy Office (Belspo);
Germany {\textendash} Bundesministerium f{\"u}r Forschung, Technologie und Raumfahrt (BMFTR),
Deutsche Forschungsgemeinschaft (DFG),
Helmholtz Alliance for Astroparticle Physics (HAP),
Initiative and Networking Fund of the Helmholtz Association,
Deutsches Elektronen Synchrotron (DESY),
and High Performance Computing cluster of the RWTH Aachen;
Sweden {\textendash} Swedish Research Council,
Swedish Polar Research Secretariat,
Swedish National Infrastructure for Computing (SNIC),
and Knut and Alice Wallenberg Foundation;
European Union {\textendash} EGI Advanced Computing for research;
Australia {\textendash} Australian Research Council;
Canada {\textendash} Natural Sciences and Engineering Research Council of Canada,
Calcul Qu{\'e}bec, Compute Ontario, Canada Foundation for Innovation, WestGrid, and Digital Research Alliance of Canada;
Denmark {\textendash} Villum Fonden, Carlsberg Foundation, and European Commission;
New Zealand {\textendash} Marsden Fund;
Japan {\textendash} Japan Society for Promotion of Science (JSPS)
and Institute for Global Prominent Research (IGPR) of Chiba University;
Korea {\textendash} National Research Foundation of Korea (NRF);
Switzerland {\textendash} Swiss National Science Foundation (SNSF).

\end{document}